\documentclass[fleqn,usenatbib]{mnras}

\usepackage{newtxtext,newtxmath}

\usepackage[T1]{fontenc}
\usepackage{ae,aecompl}

\usepackage{graphicx}	
\usepackage{amsmath}	
\usepackage{amssymb}	
\usepackage{hyperref}

\def\e{\boldsymbol{e}}

\def\bnu{\boldsymbol{\nu}}
\def\btheta{\boldsymbol{\theta}}

\def\S{\boldsymbol{S}}

\title[Time-domain filtering for improving calibration in the presence of mutual coupling]{Mitigating calibration errors from mutual coupling with time-domain filtering of 21 cm cosmological radio observations}

\author[Charles et al.]
{
\parbox{\textwidth}{ N. Charles$^{1}$ \thanks{ntsikelelo.charles@nwu.ac.za},
N. S. Kern$^{2,\dagger}$,
R. Pascua$^{3}$,
G. Bernardi$^{5,6,4}$,
L. Bester$^{4,6}$,
O. Smirnov$^{6,4}$,
E. d. L. Acedo$^{7}$,
Z. Abdurashidova$^{8}$,
T. Adams$^{4}$,
J. E. Aguirre$^{10}$,
R. Baartman$^{4}$,
A. P. Beardsley$^{11}$,
L. M. Berkhout$^{12}$,
T. S. Billings$^{10}$,
J. D. Bowman$^{12}$,
P. Bull$^{13}$,
J. Burba$^{13}$,
R. Byrne$^{14}$,
S. Carey$^{7}$,
K. Chen$^{16,2}$,
S. Choudhuri$^{17}$,
T. Cox$^{8}$,
D. R. DeBoer$^{18}$,
M. Dexter$^{18}$,
J. S. Dillon$^{8}$,
S. Dynes$^{16}$,
N. Eksteen$^{4}$,
J. Ely$^{7}$,
A. Ewall-Wice$^{19}$,
R. Fritz$^{4}$,
S. R. Furlanetto$^{20}$,
K. Gale-Sides$^{7}$,
H. Garsden$^{13}$,
B. K. Gehlot$^{12}$,
A. Ghosh$^{9}$,
A. Gorce$^{21}$,
D. Gorthi$^{8}$,
Z. Halday$^{4}$,
B. J. Hazelton$^{14,15}$,
J. N. Hewitt$^{16,2}$,
J. Hickish$^{18}$,
T. Huang$^{7}$,
D. C. Jacobs$^{12}$,
A. Josaitis$^{7}$,
J. Kerrigan$^{22}$,
P. Kittiwisit$^{9}$,
M. Kolopanis$^{12}$,
A. Lanman$^{22}$,
A. Liu$^{3}$,
Y.-Z. Ma$^{25}$,
D. H. E. MacMahon$^{18}$,
L. Malan$^{4}$,
K. Malgas$^{4}$,
C. Malgas$^{4}$,
B. Marero$^{4}$,
Z. E. Martinot$^{10}$,
L. McBride$^{21}$,
A. Mesinger$^{26}$,
N. Mohamed-Hinds$^{14}$,
M. Molewa$^{4}$,
M. F. Morales$^{14}$,
S. Murray$^{26}$,
B. Nikolic$^{7}$,
H. Nuwegeld$^{4}$,
A. R. Parsons$^{8}$,
N. Patra$^{8}$,
P. L. Plante$^{23,24}$,
Y. Qin$^{27}$,
E. Rath$^{16,2}$,
N. Razavi-Ghods$^{7}$,
D. Riley$^{16}$,
J. Robnett$^{28}$,
K. Rosie$^{29}$,
M. G. Santos$^{4,9}$,
P. Sims$^{12}$,
S. Singh$^{3}$,
D. Storer$^{14}$,
H. Swarts$^{4}$,
J. Tan$^{10}$,
M. J. Wilensky$^{13}$,
P. K. G. Williams$^{30,31}$,
P. v. Wyngaarden$^{4}$,
H. Zheng$^{2}$}
\vspace{0.4cm} \\
\parbox{\textwidth}{
$^{1}$ Centre for Space Research, North-West University, 11 Hoffmann Street,\\
$^{2}$ Department of Physics, Massachusetts Institute of Technology, Cambridge, MA\\
$^{3}$ Department of Physics and McGill Space Institute, McGill University, 3600 University Street, Montreal, QC H3A 2T8, Canada\\
$^{4}$ South African Radio Astronomy Observatory, Black River Park, 2 Fir Street, Observatory, Cape Town, 7925, South Africa\\
$^{5}$ INAF-Istituto di Radioastronomia, via Gobetti 101, 40129 Bologna, Italy\\
$^{6}$ Department of Physics and Electronics, Rhodes University, PO Box 94, Grahamstown, 6140, South Africa\\
$^{7}$ Cavendish Astrophysics, University of Cambridge, Cambridge, UK\\
$^{8}$ Department of Astronomy, University of California, Berkeley, CA\\
$^{9}$ Department of Physics and Astronomy,  University of Western Cape, Cape Town, 7535, South Africa\\
$^{10}$ Department of Physics and Astronomy, University of Pennsylvania, Philadelphia, PA\\
$^{11}$ Department of Physics, Winona State University, Winona, MN\\
$^{12}$ School of Earth and Space Exploration, Arizona State University, Tempe, AZ\\
$^{13}$ Jodrell Bank Centre for Astrophysics, University of Manchester, Manchester, M13 9PL, United Kingdom\\
$^{14}$ Department of Physics, University of Washington, Seattle, WA\\
$^{15}$ eScience Institute, University of Washington, Seattle, WA\\
$^{16}$ MIT Kavli Institute, Massachusetts Institute of Technology, Cambridge, MA\\
$^{17}$ Centre for Strings, Gravitation and Cosmology, Department of Physics, Indian Institute of Technology Madras, Chennai 600036, India\\
$^{18}$ Radio Astronomy Lab, University of California, Berkeley, CA\\
$^{19}$ Department of Physics, University of California, Berkeley, CA\\
$^{20}$ Department of Physics and Astronomy, University of California, Los Angeles, CA\\
$^{21}$ Institut d’Astrophysique Spatiale, CNRS, Université Paris-Saclay, 91405 Orsay, France\\
$^{22}$ Department of Physics, Brown University, Providence, RI\\
$^{23}$ Department of Computer Science, University of Nevada, Las Vegas, NV\\
$^{24}$ Nevada Center for Astrophysics, University of Nevada, Las Vegas, NV\\
$^{25}$ Department of Physics, Stellenbosch University, Stellenbosch, South Africa\\
$^{26}$ Scuola Normale Superiore, 56126 Pisa, PI, Italy\\
$^{27}$ Research School of Astronomy and Astrophysics, Australian National University, Canberra, ACT 2611, Australia\\
$^{28}$ National Radio Astronomy Observatory, Socorro, NM 87801, USA\\
$^{29}$ SAAO\\
$^{30}$ Center for Astrophysics, Harvard \& Smithsonian, Cambridge, MA\\
$^{31}$ American Astronomical Society, Washington, DC\\
$^{\dagger}$ NASA Hubble Fellow\\
}
}

\date{Accepted XXX. Received YYY; in original form ZZZ}

\pubyear{2022}

\begin{document}
\label{firstpage}
\pagerange{\pageref{firstpage}--\pageref{lastpage}}
\maketitle

\begin{abstract}
The 21~cm transition from neutral Hydrogen promises to be the best observational probe of the Epoch of Reionisation (EoR). This has led to the construction of low-frequency radio interferometric arrays, such as the Hydrogen Epoch of Reionization Array (HERA), aimed at systematically mapping this emission for the first time. Precision calibration, however, is a requirement in 21 cm radio observations. Due to the spatial compactness of HERA, the array is prone to the effects of mutual coupling, which inevitably lead to non-smooth calibration errors that contaminate the data. When unsmooth gains are used in calibration, intrinsically spectrally-smooth foreground emission begins to contaminate the data in a way that can prohibit a clean detection of the cosmological EoR signal. In this paper, we show that the effects of mutual coupling on calibration quality can be reduced by applying custom time-domain filters to the data prior to calibration. We find that more robust calibration solutions are derived when filtering in this way, which reduces the observed foreground power leakage. Specifically, we find a reduction of foreground power leakage by 2 orders of magnitude at $k_\parallel \approx 0.5$~h~Mpc$^{-1}$.

\end{abstract}

\begin{keywords}
cosmology: dark ages, reionization, first stars -- cosmology: observations -- methods: data analysis -- instrumentation: interferometers
\end{keywords}



\section{Introduction}
The detection of the redshifted 21~cm emission line from neutral Hydrogen during the Epoch of Reionisation (EoR) is one of the main goals of (current and upcoming) low-frequency radio telescopes such as the Low-Frequency Array \citep[LOFAR,][]{VanHaarlem2013}, the Murchison Widefield Array \citep[MWA,][]{Tingay2013}, the Giant Metrewave Radio Telescope EoR experiment \citep[GMRT,][]{Paciga2013}, the Hydrogen Epoch of Reionisation Array \citep[HERA,][]{DeBoer2017}, and the Square Kilometre Array \citep[SKA,][]{Koopmans2015}. The EoR is one of the least constrained areas of cosmology, with several cosmological models predicting different reionisation scenarios. Thus, more sensitive power spectrum limits are still required to discriminate between reionisation models. The achievement of sensitive power spectrum limits will be challenged by the presence of systematic errors, either caused by instrument response \citep[e.g.,][]{Barry2016, Fagnoni2019, Charles2022, Kim2022} or by calibration errors due to the use of incomplete sky models during calibration \citep{Wijnholds2016, Kern2020a, Charles2023}.

Measurements of the 21~cm signal are challenged by the presence of foreground emission from the Galaxy and extra-galactic sources, which are orders of magnitude brighter \citep[e.g.,][]{Santos2005, Bernardi2009, Ali2015}. Fortunately, foreground emission is spectrally smooth, unlike the 21~cm emission line, which fluctuates rapidly \citep[e.g.,][]{Santos2005}. The delay spectrum approach is a method designed to separate the 21~cm signal from the foreground emission, making use of interferometric delays to isolate the power spectrum of the 21~cm emission \citep{Parsons2012a, Liu2020}. Due to the unsmooth spectral nature of the 21~cm signal, its power spectrum appears at all $k$ modes. In contrast, the foreground emission, due to its smooth spectral emission, is limited to a wedge-like region in $k-$space \citep{Thyagarajan2013, Pober2013a, Liu2014a, Liu2014b, Barry2016, Patil2016, Ewall-Wice2017, Orosz2019, Morales2019, Byrne2019, Kern2020a, Dillon2020, Byrne2021}. The region in $k$ space where the 21~cm signal is isolated from the foreground emission is referred to as the \textit{EoR window}. 

However, the clean separation of the 21~cm signal requires the precision calibration of interferometric data \citep{DeBoer2017, Berkhout2024}. Closely packed radio interferometric arrays, like HERA, are prone to effects of mutual coupling, where the sky signal incident on an antenna is partially reflected and subsequently received by a nearby antenna. Mutual coupling in a multi-element array such as HERA is complex, and it is difficult to fully model and characterise its impact directly \citep[e.g.,][]{Sutinjo2015, Kern2020a, Fagnoni2019, Dillon2020, Josaitis2022}. If the effects of mutual coupling are not accounted for during calibration, they ultimately lead to unsmooth calibration errors, and 
cause foreground emission to leak into the EoR window \citep{Kern2020b, Orosz2019, Josaitis2022}.  

Custom time-domain filters, called \emph{fringe rate filters}, have been widely used to partially mitigate these systematics in drift-scan radio observations as a post processing step \citep{Parsons2009, Parsons2016, Ali2015, Kolopanis2019, Kern2020a, HERA2022a, HERA2023, Garsden2024}. Recent works have also explored how these filters can be used to mitigate calibration errors from inaccurate sky models \citep{Charles2023}, and from antenna displacement errors \citep{Kim2022, Kim2023}.
However, previous works have not explored how these filters could be used to mitigate mutual coupling systematics in the raw data prior to calibration in order to improve calibration quality.

In this work, we investigate the prospect of improving HERA's calibration pipeline by filtering some of the mutual coupling features in the data before running calibration, in order to mitigate their impact on the recovered antenna gain solutions. We aim to do this through the application of fringe rate filters \citep{Parsons2009}. We consider two specific kinds of filters: a simple baseline-independent high-pass filter, which we refer to as the notch filter, and a baseline-dependent filter, referred to as the main-lobe filter \citep{Charles2023, Garsden2024, Rath2024}. We assess the calibration improvement by computing the reduced chi-squared after calibration and by further examining the structure of the recovered gains in Fourier space to assess the amount of residual spectral structure. Lastly, we show how our technique helps mitigate the foreground leakage in the EoR window. Note that the exact form of the adopted main-lobe filter is specific to this work: other works may use similar kinds of "main-lobe filters" but may have slightly different parameterizations depending on their desired outcome.

This paper is divided into three sections, in Section~\ref{sec:The_Calibration_Problem} we give an overview of calibration and the simulation set up, Section \ref{sec:application_of_fringe_filters}  we discuss the application of fringe rate filters before calibration and the relevant metrics that we use to measure the improvement in calibration, and we conclude in Section~\ref{sec:conclusion}. 
 
\section{Calibration overview}
\label{sec:The_Calibration_Problem}

\subsection{Sky-model based calibration}

When observing the sky with a radio interferometer, the electromagnetic waves from celestial sources are measured by two antennas that are correlated, forming interferometric visibilities.
The measured complex-valued visibilities are related back to the intrinsic sky emission and the instrumental response of the antennas via the \emph{radio interferometer measurement equation} \citep[RIME;][]{Hamaker1996, Smirnov2011}.
\begin{equation}
\label{eq:rime}
    V^0_{ij}(\nu)=\iint J(\boldsymbol{s},\nu))\, I(\boldsymbol{s}, \nu) \, \ J^{*}(\boldsymbol{s},\nu) \, e^{-2\pi \imath \frac{\nu}{c} \boldsymbol{b}_{ij} \cdot \boldsymbol{s}} \, \, \frac{dldm}{n(\boldsymbol{s})},
\end{equation}
where $V^0_{ij}$ is the visibility formed between antenna $i$ and $j$, and $^*$ denotes the complex conjugate, $J_i(\boldsymbol{s},\nu)$ denotes Jones terms describing the propagation effects along the path from the source to the $i-$th antenna in the array. $I(\boldsymbol{s},\nu)$ is the brightness distribution of the sky, $\nu$ is the frequency of the incoming electromagnetic wave, $\boldsymbol{b}_{ij}$ is a baseline vector connecting antenna $i$ and $j$, $\boldsymbol{s} = [l, m, n]^T$ is the position vector on the celestial sphere with an origin centred at the target field (phase centre), and $(l, m, n)$ are the direction cosines, with $n = \sqrt{1 - l^2 - m^2}$. Note that in this paper, we will only consider a single polarization. The Jones terms are classified into two main groups: direction-dependent and direction-independent. The direction-dependent Jones varies with observing direction, these effects include the primary beam $E(s,\nu)$ response of the antenna, which varies across the sky for a fixed pointing. Whilst the direction-independent Jones terms are independent of observing direction. These effects include (not limited to) the band-pass response of the antenna $B(\nu)$ and gain amplification term $G(\nu)$.  

Note that in this paper, Equation~\ref{eq:rime} assumes a trivial primary beam response in that the signal from the celestial source is completely absorbed by array antennas. However, in real observations, a portion of the signal is reflected due to departure from the conjugate match at the terminals of antenna feeds. If the array is compact, the reflected signal will be absorbed by a neighbouring antenna. This effect, known as \textit{mutual coupling}, is a known concern for interferometric 21~cm experiments. More analysis of the upper limits of the power spectrum cite mutual coupling as one of the unresolved possibilities for their excess power \citep{Ali2015, Ewall-Wice2016a, HERA2022a}. 

Recent works have developed semi-analytic approaches to modelling the complex nature of coupling effects within an interferometric array.
\citet{Kern2019, Kern2020a} developed a semi-analytic approach for single antenna-pair coupling, which was highly effective in suppressing observed coupling in HERA Phase I data \citep{HERA2022a, HERA2023}.
Future work expanded upon this approach for multi-antenna coupling, which showed new phenomenology that matched the observed systematics in HERA Phase II data \citep{Josaitis2022, Rath2024}.

In this work, we will consider only first-order, multi-antenna mutual coupling effects as described in \citet{Josaitis2022}, where each antenna in the array absorbs most of the incident signal from the source, and some of the signal is reflected and completely absorbed by other antennae in the array (i.e., the incident signal is only reflected once). If we consider unpolarised radiation from the sky and assume that all beams in the array have identical electromagnetic properties, that is they have the same reflection and absorption coefficients, then the first-order visibilities (visibilities with first-order mutual coupling effects) are given by: 

\begin{equation}
\label{eq:firs_order_vis} 
    V^{1}_{ij} =  V^{0}_{ij} + \bigg(\sum_{k \neq i}\, V_{ik}^{0} X^*_{jk}\, + X_{ik} V_{kj}^{0} \ \bigg),
\end{equation} 
where
\begin{equation}
   X_{ik}= \frac{i \eta_0}{4\lambda} \frac{\Gamma}{R \, |\boldsymbol{b}_{ij}|} \, e^{2\pi i \frac{\nu}{c} |\boldsymbol{b}_{ij}| } \, h_0^2 \, J(b_{ik}) J^*(b_{ki}) 
\end{equation}

where $\eta_0$ is the impedance of free space, $\Gamma$ is the frequency-dependent reflection coefficient, $R$ is the real part of the impedance term and $ J(b_{ki}) $ denotes direction-dependent effects (primary beam response) along the path travelled by the reflected signal from antenna $k$ to antenna $i$ and $h_0$ is the antennas so-called effective height \citep{Josaitis2022, Rath2024}.

In real observations, antennas in an array also impart direction-independent corruptions to the visibilities, which can be modelled using antenna-based complex gain terms. The corrupted visibilities are related to intrinsic visibilities as
\begin{equation}
    V^{c}_{ij}=g_i\, V_{ij}\, g_j^*\,  + n_{ij},
    \label{eq:corrup_model}
\end{equation}
where $V^{c}_{ij}$ are the corrupted (or measured) visibilities, $g_i$ and $g_j$ are the complex gain terms of antenna $i$ and $j$, and $n_{ij}$ is the complex thermal noise generated by the telescope.

Antenna gain calibration is the process of deriving the direction-independent gain terms ($g_i$) from the data and then correcting the data to remove their effects.
This is commonly done by constructing a model of the true visibilities and then minimising a $\chi^2$ statistic:
\begin{equation}
    \chi^2 =\sum_{i,j} \frac{ |V_{ij}^{c}-g_i\, g_j^*V^{m}_{ij}|^2}{\sigma_{ij}^2},
\end{equation}
where $V_{ij}^{m}$ are the constructed model visibilities, and $\sigma_{ij}^2$ is the variance of the visibility, and the sum runs over all unique antenna pairs.

The chi-square minimisation depends on the accurate modelling of the intrinsic sky visibilities $V_{ij}$. It is crucial, therefore, that the sky model matches the intrinsic sky brightness as closely as possible. Unmodelled mutual coupling effects can cause significant deviation between intrinsic sky visibilities and the model visibilities in calibration, leading to calibration errors that can impart spectral structure on smooth foregrounds \citep{Fagnoni2019, Kern2020a}.

\subsection{Redundant calibration}

Redundant calibration is less reliant on a sky model compared to sky-base calibration, to solve for antenna gains, the gains are obtained by exploiting the internal redundancy of the interferometric array. Redundant baselines have the same length and orientation and, therefore, measure the same Fourier mode of the sky brightness distribution.
For example, assume we have a single baseline type, $A$, uniquely identified via its baseline vector $\boldsymbol{b}_A$. For all antenna pairs $i,j$ that share this baseline vector, the calibration equation, Equation \ref{eq:corrup_model} from before now becomes
\begin{equation}
       V^{c}_{ij} = g_i \, V_{A} \, g_j^* + n_{ij}, 
\label{eq:red_cal}
\end{equation}
where $V_{A}$ is now a parameter of the model, called the ``redundant model visibility''.
This is repeated for all unique baseline types, eventually building up an overconstrained system of equations, which can be solved via a $\chi^2$ minimization \citep[e.g.,][]{Liu2010}:
\begin{equation}
    \chi^2 =\sum_{i,j} \frac{ |V_{ij}^{c}-g_i\, g_j^*V_{A}|^2}{\sigma_{ij}^2},
\end{equation}
where $V_{A}$ is the corresponding redundant model visibility for the $\boldsymbol{b}_{ij}$ baseline vector. Note that in this paper, the $\chi^2$ is both a function of frequency and time. The gain solutions obtained from the system of linear equations set up in the redundant calibration are not unique; there are at least four degenerate parameters that need to be solved for after redundant calibration: the overall gain amplitude, the model visibility amplitude and the phase gradient across the array in the east-west and north-south orientations \citep{Liu2010, Zheng2014, Dillon2016, Dillon2018, Liu2020}. These parameters are, in principle, a function of polarization, frequency and time and require a sky model to be constrained.  
It is in this step that, redundant calibration itself makes use of a sky model. In this study, we will refer to this step in calibration as \text{absolute calibration} to differentiate this from \text{sky calibration} or the typical sky-based calibration discussed previously.

In the case of perfect calibration, where the sky model is complete, model visibilities only differ from the data by noise. The expectation value of chi-square $\langle \chi^2 \rangle$ is thus two times the degrees of freedom (DoF) for complex data \citep{Dillon2020}. The number of DoF, in general, is given by the difference between the number of data points and the number of fitted parameters. If we consider a single polarization and the case of sky-based calibration, the DoF are given by the difference between the number of visibilities and the number of antennas:
\begin{equation}
    {\rm DoF} = \frac{N(N-1)}{2} - N = N_{\rm bl} - N,
\end{equation}
where $N$ is the number of antennas and $N_{\rm bl}$ is the number of baselines given $N$ antennas. If we histogram the $\chi^2/{\rm DoF}$ quantity, it is expected to follow a theoretical chi-square distribution \citep{Dillon2020}. 
In the case of redundant calibration, the DoF is given by:
\begin{equation}
    {\rm DoF} = N_{\rm bl}-N_{\rm ubl}-N+2,
\end{equation}
where $N_{\rm ubl}$ is the number of unique baselines that have a single sky model visibility $V_A$, i.e. the number of redundant baseline groups \citep{Dillon2020}.

\subsection{Simulation set up}

In this paper, we simulate sky models composed of diffuse emission from GSM \citep{Oliveira2008} and compact source flux density consistent with the GLEAM catalogue \citep{Hurley-Walker2017}, but with no spatial clustering of sources (uniformly distributed). We also include bright radio galaxies \citep[Fornax~A, Hydra~A, Pictor~A, Hercules~A, Virgo~A, Cygnus~A, Cassiopeia~A;][]{MCKinley2015, Byrne2022}, modelled as compact sources. To simulate mutual coupling effects, we make use of the mutual coupling simulation formalism from \citet{Josaitis2022} to produce visibility products with first order mutual coupling effects. Their formalism included modelling the first-order antenna-antenna coupling in radio interferometers.  Specifically, they derive a semi-analytic model of the interferometric visibility Equation~\ref{eq:firs_order_vis}, in which first-order coupling effects are considered. In their model, incident radiation enters all interferometric elements in the array, and then each element absorbs most of the incident radiation, but some of the power is then reflected due to departure from the conjugate match at the terminals of the antenna feeds. Thus, each element in the array not only absorbs the incident array from the sky but re-radiates it across the array, which is subsequently absorbed by other elements. This simulation only considers the first-order formalism, and thus, it only considers the effects of re-radiated radiation being absorbed only once by all the array elements. In this paper, we refer to simulated visibilities containing effects of mutual coupling as first order visibilities. And simulated visibilities without effects of mutual coupling are to as zeroth order visibilities.  

\begin{figure*}
    \centering
    \includegraphics[scale=0.55]{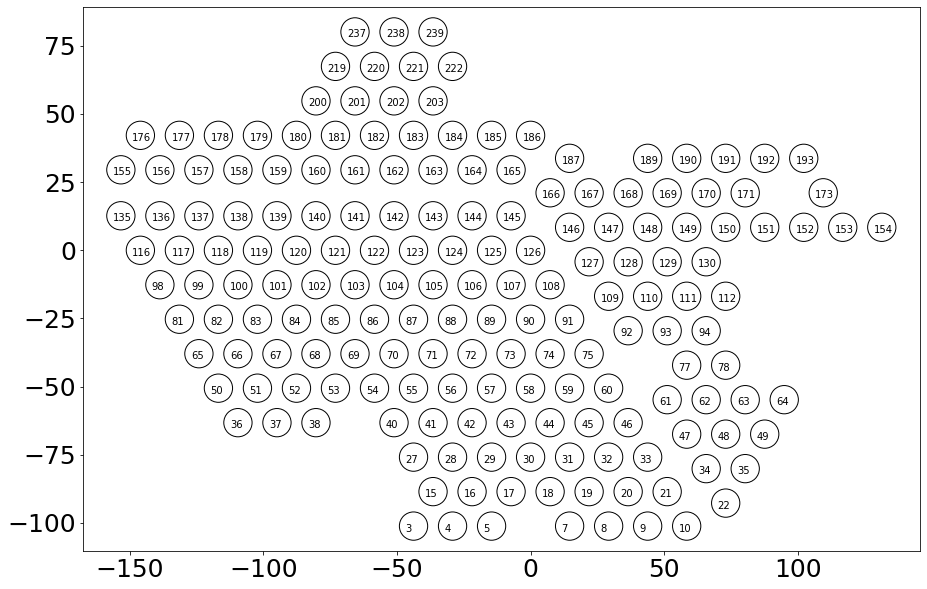}
    \caption{Simulated array layout with 174 antennas and 14.6 m
spacing between antennas.}
    \label{fig:antenna_layout}
\end{figure*}

The simulated visibilities span a frequency range of $140-190$~MHz, with a frequency resolution of $122.07$~kHz. We consider observations in the Local Sidereal Time (LST) range $0^h-4^h$, with a time resolution of $dt=58$~s. The simulation is done for a HERA array consisting of 174 antennas shown in Figure~\ref{fig:antenna_layout}. 
Figure~\ref{fig:non_redundant_vis} shows the visibility spectra from 14~m East-West redundant baselines with (first order visibilities) and without mutual coupling (zeroth order visibilities). The visibility spectra from redundant baselines have a higher amplitude at low frequency, as expected, due to the brightness of diffuse emission at low frequencies \citep{Bernardi2010} for both zeroth and first order visibility. Primary beam models containing the effects of mutual coupling differ from the primary beam model that is without mutual coupling \citep{Fagnoni2019}. As a result, the visibility spectra from the 14~m baselines without mutual coupling (solid red line) differ significantly from the visibility spectra of the same baseline with mutual coupling. Also, notably, the visibility spectra from 14~m redundant baselines with mutual coupling deviate from each other, and these deviations are as high as 6\% across the frequency band.

We make use of gains that are similar to ones simulated by \citet{Charles2023} to create mock raw data.  The gain $g$ for the $j$th antenna is given by:
\begin{equation}
    g_j(\nu)=A_j (\nu) e^{i \phi_j(\nu)},
\end{equation}
 where the amplitude $A$ follows a frequency power law 
\begin{equation}
    A_j(\nu)=A_j\,  \left( \frac{\nu}{ {\rm150~MHz}} \right)^{b_j},
\end{equation}
and $A_j$ and $b_j$ are drawn for each antenna from a Gaussian distribution $\mathcal{N} \sim (\bar{x}_{A}=0.30, \sigma_{A}=0.001)$ and $\mathcal{N} \sim (\bar{x}_b=-2.6,\sigma_b=0.2)$, respectively. The phase $\phi_i(\nu)$ is modeled as:
\begin{equation}
  \phi_j(\nu)=\sin{(w_a \, \nu)} + \cos{(w_b \, \nu)},
\end{equation}
where $w_a$ and $w_b$ are drawn for each antenna from a Gaussian distribution $ \mathcal{N} \sim (\bar{x}_w=0.005, \sigma_w=0.0005)$. The mean values of the parameters $A$ and $b$, as well as their standard deviation $\sigma_A$ and $\sigma_b$, are informed by the gain solutions from actual HERA observations \citep{Kern2020a}. The mean phase variation $\bar{w}$ is based on single antenna phase dependency of actual HERA gains where the cable delay and geometric phase offset have been removed, and the variation of the mean phase between antenna stations, i.e. $\sigma_w$, is chosen to be within $10\%$. We then add noise to the data, such that signal-to-noise ratio (SNR) of the simulated bandwidth $\sim 10^4$, mutual coupling effects are known to be prominent at this dynamic range \citep{Josaitis2022}. Note that this SNR is far higher than in real observations \citep{Dillon2020}.  We simulate two sets of raw data products; one includes the effects of mutual coupling, and the other does not. The latter raw data product is used to verify our pipeline and serves as a benchmark to quantify the effects of mutual coupling. 

We subdivide the band into two sub-bands, which we name low-band and high-band, with a frequency range of $140-165$~MHz and $165-190$~MHz, respectively. We make use of a sky model that contains no mutual coupling effects, i.e., our sky model is composed of zero-order visibilities only. First, we make use of redundant calibration to calibrate raw data. The gains from the redundant calibration step are then applied to the raw data, i.e., the raw data is partially calibrated. The calibrated visibilities are then further calibrated using the absolute calibration step. Here, the data is calibrated using the sky model (zeroth order visibilities), and the gains recovered from this step are combined with those recovered from the redundant calibration step to make a full set of gains. We then apply the full set of gains to calibrate the raw data. To quantify the robustness of the calibration,  we compute the reduced chi-square from redundant calibration and absolute calibration steps. 

\subsection{Calibration with mutual coupling effects}

\begin{figure*}
    \centering
    \begin{tabular}{cc}
       \includegraphics[scale=0.6]{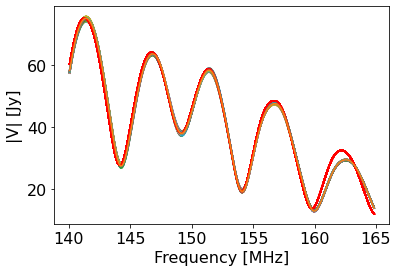}  &  \includegraphics[scale=0.6]{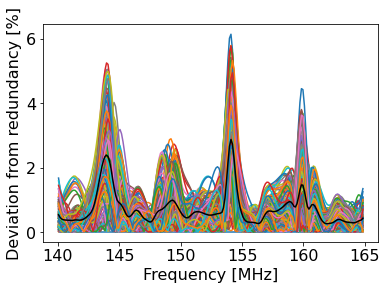}
         
    \end{tabular}
    
    \caption{Left: Simulated visibility amplitude of from 14~m East-West redundant baselines at LST$=0.20^h$. The red line denotes the visibility spectra from the data set without mutual coupling. The different colours denote the visibility spectra from 14 m redundant baselines with mutual coupling. Right: Shows the percentage deviation of visibility spectra amplitude from the mean visibility spectra amplitude of 14~m redundant baselines with mutual coupling. The black line shows the average percentage difference over all baselines within the 14~m redundant. The amplitude visibility spectra of the redundant baseline differ from each other (and the mean) due to the effects of mutual coupling. }
    \label{fig:non_redundant_vis}
\end{figure*}

\begin{figure}
    \centering
    \includegraphics[scale=0.6]{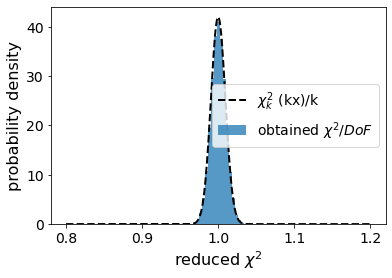}
    \caption{Histogram of the reduced $\chi^2$ for visibility data without mutual coupling (blue) for low band, i.e. $\mathrm{\chi^2/DoF}$. The dashed line is the theoretical reduced $\chi^2$-distribution with $k = 2 \times \mathrm{DoF}$, i.e. the chi-square distribution DoF, and $x=\mathrm{\chi^2/DoF}$. This is a simple demonstration that our calibration pipeline works as expected when the effects of mutual coupling are excluded.}
    \label{fig:redundant_chis_high}
\end{figure}

\begin{figure*}
        \centering
        \begin{tabular}{cc}
             \includegraphics[scale=0.6]{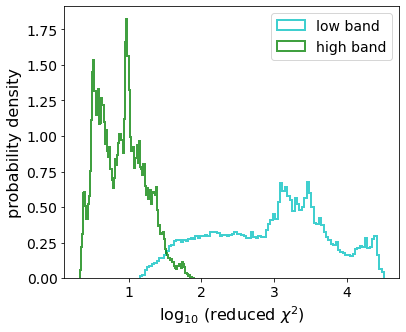} & \includegraphics[scale=0.6]{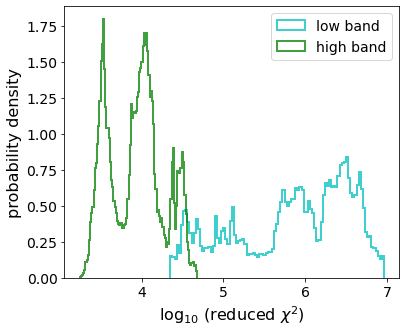} 
            
        \end{tabular}
        \caption{Left: Histogram of the reduced $\chi^2$ recovered from redundant calibration step for both low sub-band (cyan) and high sub-band (green). Right: same but for chi-square in absolute calibration step calibration.}
        \label{fig:non_redundant_chis_low}
\end{figure*}

\begin{figure}
    \centering
    \includegraphics[scale=0.6]{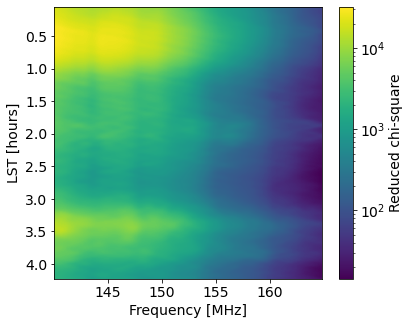}
    \caption{Reduced chi-square from redundant calibration for the data set with mutual coupling as a function of both frequency and LST. All LSTs demonstrate poor data-to-model fit after calibration due to the presence of mutual coupling. The spike in chi-square at LST=$3.3^h$ is likely driven by Fornax A, and at LST=$0.5^h$ driven by the Galactic plane entering the primary beam sidelobes.}
    \label{fig:non_redundant_chis_low_as_function_LST}
\end{figure}

Figure~\ref {fig:redundant_chis_high} shows the reduced chi-square obtained from the redundant calibration step for an ideal calibration scenario, i.e., where the raw data does not contain the effects of mutual coupling. The histogrammed reduced chi-square from the data set without mutual coupling, as expected, follows the theoretical $\chi^2$ distribution, indicating that model and raw data visibilities only differ by noise, and the recovered gains from calibration are perfect (match the true gains), thus also verifying the calibration pipeline works as expected. Figure~\ref {fig:non_redundant_chis_low} shows the recovered chi-square from the data set with mutual coupling for both low and high sub-bands. The chi-square is notably biased to higher values for both sub-bands, but notably significantly higher at the low sub-band. The mean value of the reduced chi-square $\chi^2 /\mathrm{DoF}$ is $3.3\times 10^3$ and $1.1\times 10^3$ for low and high sub-band, respectively. This bias is due to differences in visibility products from redundant baselines (see Figure~\ref{fig:non_redundant_vis}) caused by the effects of mutual coupling, since redundant calibration makes no use of the sky model. Given the mean value of the reduced chi-square, we can also infer that the redundant calibration errors are larger at lower frequencies (low band) than at high frequencies. The reason for this is that the increased brightness at lower frequencies of the diffuse emission, which couples to the primary beam response, thereby effectively enhancing differences in the apparent sky observed by two antennas in the array. Thus, when redundantly calibrating the data, larger differences are found between the visibility spectra of supposedly redundant baselines. Figure~\ref{fig:non_redundant_chis_low_as_function_LST} further shows that the reduced chi-square depends upon the looking direction, thus indicating that the chi-square depends on the brightness of the diffuse emission. 

The right panel of Figure~\ref {fig:non_redundant_chis_low} show the chi-square obtained after the absolute calibration step. The mean value of the recovered chi-square $\chi^2 /\mathrm{DoF}$ from the absolute calibration step is $1.6\times 10^6$ and $1.1\times 10^4$ for both low and high sub-band, respectively. At the low sub-band, the mean value is at least $10^2$ times greater than that obtained from the redundant calibration step. The increase in chi-square is caused by the fact in our sky model we do not incorporate the effects of mutual coupling, which are found in the data, thus, the model and raw visibility products differ significantly (see Figure~\ref{fig:non_redundant_vis}). In addition, calibration errors in the redundant calibration step are carried through to the absolute calibration step, and hence, thus calibration errors accumulate, and the obtained chi-square is significantly larger in the absolute calibration step. We note the recovered chi-square values are large compared to the values ($\chi^2 /\mathrm{DoF}$ around 1-3) one would find in typical HERA observations \citep[e.g.,][]{Dillon2020}, this is just a consequence of the low noise (high SNR) in our simulation compared to actual observations. Small differences between model and raw data visibilities are effectively 'amplified' by the small variance $\sigma^2 \sim 10^{-8}$ in our simulation, whereas in actual observation the noise variance is typically $\sigma^2 > 10^{-4}$.

\section{Fourier representation of interferometric data}
\label{sec:application_of_fringe_filters}
We will now define relevant quantities that will be used to quantify the impact of mutual coupling in calibration. Interferometric visibilities are a function of frequency ($\nu$) and time $(t)$. We refer to the Fourier dual of frequency as the delay domain, formed by taking a Fourier transform of the visibilities across the frequency axis:
\begin{equation}
\label{eq:delay_transform}
    \tilde{V}(\tau) = \int V(\nu) \, \e^{-2 \pi i \tau \nu} d\nu,
\end{equation}
where $\tau$ is a delay (in seconds) and $\tilde{V}(\tau)$ is the Fourier pair of the frequency. 

We can also define delay-transformed antenna gains to be the Fourier transform of the antenna gains along the frequency axis:
\begin{equation}
\label{eq:delay_transform_gain}
    \tilde{g}(\tau) = \int g(\nu) \, \e^{-2 \pi i \tau \nu} d\nu.
\end{equation}
We define the Fourier dual of time as the \emph{fringe rate} and the fringe-rate visibility as the Fourier transform of the visibility along the time axis:
\begin{equation}
\label{eq:fr_transform}
    \tilde{V}(f) = \int V(t) \, \e^{-2 \pi i f t} dt,
\end{equation}
where $f$ is the fringe rate (in units of Hz). For interferometric arrays that operate in drift-scan mode, the fringe rate basis also separates signals based on their relative motion through the fixed interferometric fringes, acting as another form of separation of signals on the sky \citep{Parsons2009, Parsons2016}.

\subsection{Recovered gains from calibration}
\begin{figure*}
    \centering
    \begin{tabular}{c}
         \includegraphics[scale=0.41]{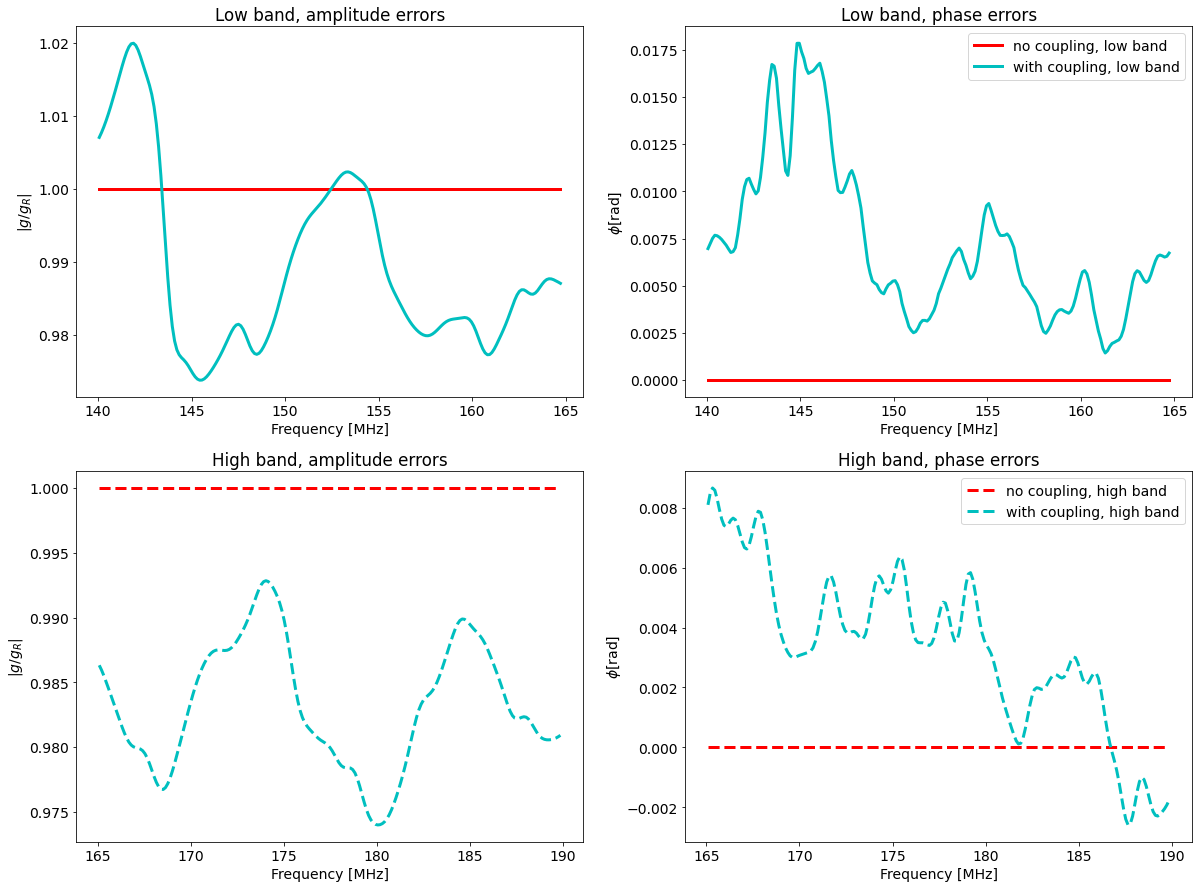}

    \end{tabular}
    
    \caption{ Top Left panel: Amplitude ratio $g/g_r$ between the gain $g$ obtained using the data set with mutual coupling and the gain $g_r$ obtained from calibrating the data set without mutual coupling average of over all antennas. Gains recovered from the absolute calibration step for both data sets with mutual coupling (cyan) and without mutual coupling (red). Top left panel: complex phase $\phi$ of the ratio $g/g_r$. Bottom panels: same but for high sub-band.  }
    \label{fig:abscal_gains_non_redundant_no_filter_low}
\end{figure*}

\begin{figure}
    \centering

         \includegraphics[scale=0.6]{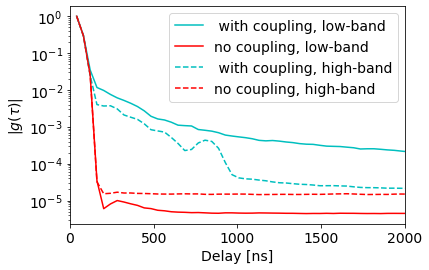}  \\
    
    \caption{Magnitude of Fourier transform of the gains along frequency i.e., delay $|g(\tau)|$ for gains recovered from absolute calibration step average of over all antennas for both data sets with mutual coupling (cyan) and without mutual coupling (red) for both low (solid line) and high (dotted line) sub-bands. }
    \label{fig:abscal_gains_no_filter_delay}
\end{figure}

Figure~\ref{fig:abscal_gains_non_redundant_no_filter_low} shows the gains recovered from absolute calibration for both low and high sub-band. Let's first consider the gains recovered from the low-band data set. The recovered gains show a pronounced frequency structure in both amplitude and phase, caused by calibration errors resulting from unmodelled mutual coupling effects. The calibration errors are unsmooth, resulting in excess power at delays $\tau>250$~ns (Figure~\ref{fig:abscal_gains_no_filter_delay}). The dynamic range, i.e., the ratio of the gain amplitude at delay $\tau=0$ to that of a specific delay $\tau$, of the gains without mutual coupling at $\tau = 250$~ns, is $\sim 10^5$. Due to the effects of mutual coupling, this is reduced to $\sim 10^2$.  Notably, the mutual coupling effects lead to excess power in the gains at a broader range of delays ($\tau>250$~ns) compared to calibration errors from incomplete sky modelling ($200 <\tau< 600$~ns) \citep[e.g.,][]{Kern2020a, Charles2023}. Indeed \citet{Josaitis2022} also found that mutual coupling effects tend to contaminate a wider area of the EoR window. The recovered gains with mutual coupling at the higher band, however, have comparatively less excess power, and the excess power spans a narrower delay range $250< \tau <2000$~ns. This is due to the reduced brightness of the diffuse emission, within the high sub-band.

\subsection{Application of Fringe Rate Filters}
\label{sub_sec:fringe_rate_filtering}

We propose the use of a fringe rate filter as a way to reduce the impact of mutual coupling in calibration. The idea is to filter out some mutual coupling effects in the raw data, thus bringing the raw data closer to the sky model. We employ a similar approach to that used by \citet{Charles2023} to improve calibration. We also apply fringe rate filtering to \emph{both} the raw data and the sky model visibilities in the same manner for each of the different filters that we use (described below). We then apply the derived gains to the unfiltered visibilities, thus mitigating concerns of a possible cosmological signal loss.

Note that, for the two filters described below, we apply them to the data using the DAYENU filtering formalism described in \citet{Ewall-Wice2020}, which relies on the Discrete Prolate Spheroidal Sequences \citep[DPSS;][]{Slepian1978}.

\subsubsection{Notch Filters}
\label{sec:notch_filters}
We first consider a symmetric baseline-independent notch filter $F(f)$ centred at $f = 0$~mHz fringe rate,  i.e. a high-pass filter, defined as 
\begin{equation}
F(f) = 
\left\{
    \begin{array}{lr}
        10^{-8}, & |f| \leq f_{\rm max}\\
              1, & |f| > f_{\rm max}
    \end{array}
\right\},
\end{equation}
with 
$f_{\rm max} = 0.25$~mHz,  we refer to the filter as $f_{25}$. See Figure~\ref{fig:Main_lobe_and_notch_filter} for a visual representation of the notch filter.

\subsubsection{Main Lobe Filters}
\label{sec:main_lobe_filters}

We also considered a second type of filter, which we refer to as a ``main lobe'' filter, because it aims to suppress emission from outside the primary beam field-of-view.
In contrast to the baseline-independent notch filter, which only filters out emission in a region near $f\sim0$~mHz, the baseline-dependent main-lobe filter suppresses the signal everywhere \emph{except} near the peak emission of the sky model in fringe rate.

Technically, the main-lobe filter is a frequency-dependent filter, but because we operate over a relatively small bandwidth (25~MHz within each band), we approximate it as frequency-independent. The response of the filter within the pre-defined fringe rate bounds is uniform, such that it can be thought of as a top-hat filter.

\begin{figure}
    \centering
    \includegraphics[scale=0.6]{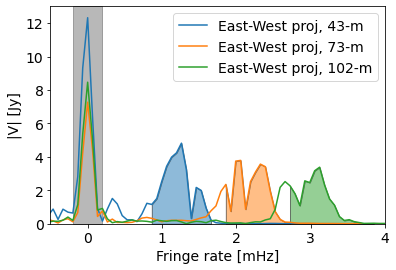}
    \caption{Visibility amplitude of the first order model visibilities in fringe rate space at a frequency $\nu=140$~MHz for three baselines with an East-West projection of 43~m (blue), 73~m (orange) and 102~m (green). The colour field regions denote the sky emission in fringe rate space that is retained for calibration, when the main lobe filter is used. Also shown in grey is the region around zero fringe rate, where the sky emission is filtered upon application of the notch filter.}
    \label{fig:Main_lobe_and_notch_filter}
\end{figure}

The bounds of the baseline-dependent main-lobe filter are determined by its centre $f_0$  and its half-width $f_w$. These parameters are determined for each baseline by fitting a Gaussian profile $G(f)$ to the sky model visibilities in fringe rate space:
\begin{equation}
    G(f) = A \, e^{-\frac{(f - f_{0})^2}{2 \sigma^2}},
\end{equation}
where $A$ is the amplitude of the Gaussian, $\sigma$ is its standard deviation, and $f_0$ is its mean.
After the fit, we set the main lobe filter centre to be $f_0$ and its half-width to be $f_w = 2\sigma$, such that the full width of the main lobe filter is $4\times$ the fitted Gaussian's standard deviation. We denote the baseline dependent main lobe filter as $M_{de}$ (Figure~\ref{fig:Main_lobe_and_notch_filter}).

\subsection{Using the Notch Filter to Improve Calibration}
In Figure~\ref{fig:non_redundant_chis_low_as_function_LST}, we have noted that the recovered chi-square is dependent on the looking direction (LST), we have shown that the chi-square is coupled to the diffuse emission, that is, it is higher when the diffuse emission is bright. To show the effectiveness of the fringe rate filters we consider a field where the chi-square is largest (bright diffuse emission), i.e., we are considering the 'worst case-scenario'. We consider a field centred at an LST range $0^h-1^h$. We first examine the $\chi^2$ obtained after the absolute sky calibration step. The sky emission can be separated in fringe space, as different parts of the sky occupy different fringe rate values, for an array observing in tracking mode i.e., with the sky emission drifting across the array \citep{Parsons2016}. For a baseline in the HERA array, we note, especially, that emission originating from the celestial South Pole will occupy fringe rate value zero, as sources at the celestial South Pole do not drift across the array \citep{Parsons2016}. Thus, we expect these sources to fall on primary beam side lobes. Due to the unsmooth spectra of the primary beam with mutual coupling \citep{Fagnoni2019}, when this emission couples with primary beam response, we expect the resulting calibration errors to be unsmooth, as shown in Figure~\ref{fig:abscal_gains_non_redundant_no_filter_low}. Note, however, that this is only true for a baseline with a non-zero East-West projection \citep{Parsons2016}. \citet{Kern2020b} also showed that mutual coupling effects occupy in fringe rate space occupy fringe rate values near zero. Thus, in this paper, we design fringe rate filters to filter out mutual coupling effects at zero fringe rate. We employ a baseline cut and downweight all baselines during the calibration process that have an East-West projection of less than 30~m, to ensure that all baselines have a peak of the main lobe emission (in fringe rate space) at fringe rate values greater than $0.5$~mHz.  

\begin{figure}
    \centering
    \begin{tabular}{cc}
        \includegraphics[scale=0.6]{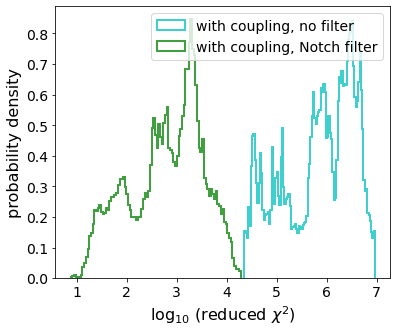} &
         
    \end{tabular}
    
    \caption{Histogram of the reduced chi-square recovered from the absolute calibration during the calibration of the data set with mutual coupling with no filter (cyan) and where we apply the notch fringe rate filter $f_{25}$ before calibration green). The reduced $\chi^2$ improves significantly after using a notch filter, in terms of its peak value being closer to one.}
    \label{fig:chis_non_redundant_no_filter_low_with_filters}
\end{figure}

\begin{figure*}
    \centering
    \begin{tabular}{cc}
         \includegraphics[scale=0.58]{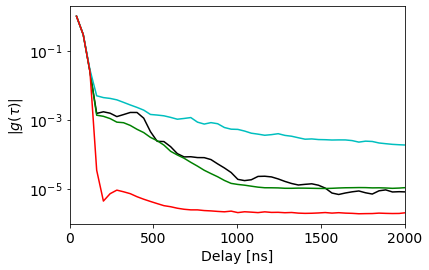} &
         \includegraphics[scale=0.58]{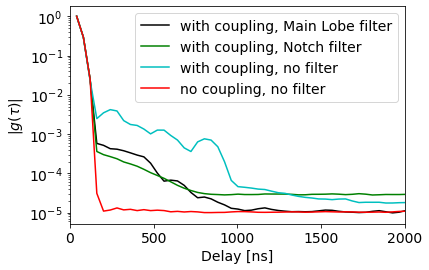}
         
    \end{tabular}
    
    \caption{Amplitude of delay transform $|g(\tau)|$ of the recovered gains after calibrating the data set with and without mutual coupling. Here, we plot the gains recovered from calibrating the data set without mutual coupling (red) and with mutual coupling (cyan) without applying filters. We then show gains after applying a notch filter $f_{25}$ (green) and the main lobe filter (black) for both low (left panel) and high (right panel) sub-band. The significant amount of spectral structure seen when calibrating the data set with mutual coupling is heavily suppressed after applying a filter, even more so when the baseline-independent notch filter is applied ($f_{25}$). }
    \label{fig:abscal_gains_non_redundant_no_filter_low_with_filters}
\end{figure*}

Figure~\ref{fig:chis_non_redundant_no_filter_low_with_filters} shows the chi-square obtained for the low sub-band when a notch fringe filter is applied to the data set with mutual coupling before calibration. The chi-square has reduced significantly, with a mean value of $1.7\times 10^3$. This represents a reduction in $\chi^2$ of factor $\sim 10^3$. This occurs because the filter suppresses some of the mutual coupling features in the raw data, thus bringing the raw data closer to the model, leading to a smaller chi-square value. A smaller $\chi^2$ value is indicative of robust calibration. Thus, we expect the recovered gains to improve as well. Figure~\ref{fig:abscal_gains_non_redundant_no_filter_low_with_filters} shows the recovered gains after applying fringe rate filters. The recovered gains have comparatively less frequency structure in both amplitude and phase, and as a result, the delay transform of the gains has comparatively less excess power at delays $\tau >250$~ns for both the main lobe and notch filter. The dynamic range is improved by a factor $\sim~10$ at delays $\tau \approx 250$~ns for both high and low sub-band, when the notch filter $f_{25}$ is applied. However, the gains recovered after applying the main lobe filter show comparatively less improvement (compared to the notch filter), in terms of dynamic range at delays $250<\tau<1000$~ns at low sub-band. Similarly, at the high sub-band, the notch filter performs comparatively better than the main lobe filter over the delay range  $250<\tau<500$~ns. Notably, the gains recovered from the notch-filtered data have a higher noise floor compared to unfiltered data, and this is due to lower SNR as some of the sky signal is suppressed by the filter, thus reducing the SNR. However, the gains recovered when the main lobe filter is used have a lower noise floor compared to the notch filter, and this is because the main lobe filter retains most of the sky signal, and in addition, it also significantly attenuates some of the noise. Note that the current HERA calibration pipeline includes smoothing the derived gains, i.e., a prior on the spectral structure of gains is imposed, and this mitigates the some of effects of mutual coupling \citep{HERA2022a}. However, the advantage of using fringe rate filters is that, if there is real spectral structure in the intrinsic antenna gains, say above $\sim 100$~ns, then the use of fringe rate filters has a better chance of finding the true gain response.

We can see that the notch filter performs better than the main-lobe filter at lower delays $\tau<500$~ns, given the difference in the dynamic range of the recovered gains. The main key to understanding the improved performance of notch filters is to examine closely the filtering and calibration process. The main key to getting robust gains after calibration is to minimise the differences between the sky model and raw data visibility products. The main lobe filter is designed to retain the main lobe emission in fringe rate, and for a baseline with an East-West baseline projection greater than 30~m, this emission also falls within the main lobe of the primary beam, with sidelobe emission occupying fringe rate near zero. Although the effects of mutual coupling are smaller in the main lobe of the primary beam compared to the sidelobes, there are still significant differences in the antenna primary beam main lobe, between mutual coupling and non-mutual coupling beam models \citep{Fagnoni2019, Charles2022}, and these primary beam differences would contribute significantly to calibration errors, even more so when bright emission falls on the main lobe of the beam (see Figure~\ref{fig:non_redundant_chis_low_as_function_LST}). Unlike the sky emission falling on the sidelobe, the sky emission falling on the main lobe (with some mutual coupling effects) is not significantly attenuated. Thus, it dominates calibration. Mutual coupling effects in the main lobe of the beam have a much larger impact on the calibration compared to mutual coupling from the sidelobe of the primary beam, and are typically relatively smooth in frequency compared to side lobe mutual coupling effects. Thus, we would expect calibration errors due to mutual coupling effects to be relatively smooth in frequency as compared to mutual coupling effects from the sidelobes. The key goal of the filters is to suppress a large portion of emission where mutual coupling effects are pronounced, whilst retaining a good portion of the sky signal that can be used for calibration. On baselines with short East-West projection (main lobe emission is close to  $f\sim 0.5$~mHz), the baseline independent notch filter, filters a portion of the emission that lies in the main lobe emission,  whereas the main lobe filter retains this emission, which may contain some mutual coupling effects. 

\begin{figure}
    \centering
    \includegraphics[scale=0.6]{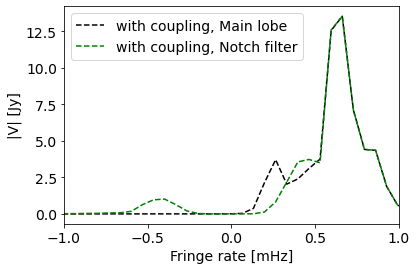}
    \caption{Simulated model visibility products of baseline (3, 42) with mutual coupling in fringe rate space, using notch filter $f_{25}$ (green) and baseline dependent main lobe filter $M_{be}$ (black). }
    \label{fig:filtered_emission_main_and_notch}
\end{figure}

\begin{figure}
    \centering
    \includegraphics[scale=0.6]{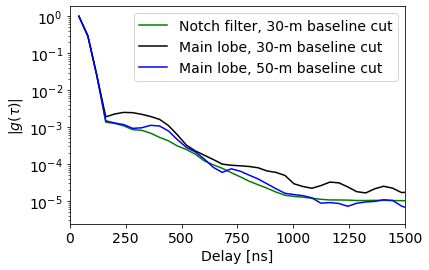}
    \caption{Delay transform of the gains $|g(\tau)|$ recovered for the absolute calibration step when the notch filter (green) and main lobe filter (black) are used, with a 30 m baseline cut. The other colours show gains recovered when the main lobe filter is used, however, the baseline cut is increased to 50 m (blue). Notably here is that the dynamic range of the gains recovered through the main lobe filter with a 50 m E-W baseline cut, has the same dynamic range ($100<\tau<250$~ns) as the notch filter with a 30 m baseline cut.}
    \label{fig:abscal_gains_filter_baseline_cut}
\end{figure}

\begin{figure}
    \centering
    \includegraphics[scale=0.6]{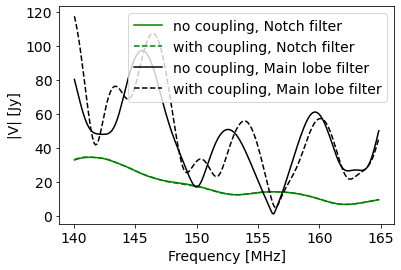}
    \caption{Visibility spectra of main lobe filtered data for both zeroth (black solid line) and first order visibilities (Black dotted line). Green is the same but for notch filtered data. Notably, the notch-filtered visibility spectra from both zeroth order and first order visibility closely match up compared to the main lobe filtered visibilities.}
    \label{fig:visibility_spectra_filtered}
\end{figure}

An example of such baseline emission is shown in Figure~\ref{fig:filtered_emission_main_and_notch}. The shoulder emission (emission at fringe rate value $f\approx 0.25$~mHz) in the main lobe is filtered out by the notch filter (green), minimising the impact of mutual coupling effects in the main lobe emission, thus bringing the raw data closer to the sky model data, reducing errors in calibration. However, the baseline-dependent main lobe filter (black) retains this emission. Figure~\ref{fig:visibility_spectra_filtered} shows the visibility spectra of the main lobe and notch filtered visibility of both zeroth order (without mutual coupling, solid line) and first order (with mutual coupling, dotted line) visibilities. The main lobe filtered visibilities (with and without mutual coupling) deviate significantly from each other as the main lobe filter essentially fails to attenuate the relatively smooth frequency varying unmodelled mutual coupling terms, which can be seen as the ripples in spectra (black dotted line). Thus, model and raw data visibilities differ significantly, leading to larger calibration errors. However, the notch-filtered visibility closely match with each other, as ripples in visibility spectra caused by mutual coupling effects are attenuated significantly, leaving an overall smooth visibility spectra for both filtered and unfiltered. The difference between filtered and unfiltered visibility spectra is notably small for the notch filter compared to the main lobe filter,  thus, the notch filter would yield smaller calibration errors, leading to more robust recovered gains that can be used to calibrate the data. In order to achieve the same improvement as the notch filter in the calibration, particularly at the low delays $\tau <500$~ns, we need to implement a E-W baseline cut larger than $50$~m. In Figure~\ref{fig:abscal_gains_filter_baseline_cut}, we show that the recovered gains from calibration using the notch filter have the same dynamic range as gains recovered from the main lobe filter, but with an E-W baseline cut of 50~m.


\subsection{Foreground Power Spectrum}

In this section, we will introduce the power spectrum as a metric to assess the impact of fringe rate filters on calibration. We compute the per-baseline power spectrum of the visibilities, $P(\tau, b)$, following the delay approximation \citep{Parsons2012a, Liu2020}.

\begin{equation}
    P(\tau, b)=|\Tilde{V}_b(\tau)|^2 \, \bigg(\frac{\lambda^2}{2k_B}\bigg)^2\bigg(\frac{D_c^2\Delta D_c}{B_{\rm eff}}\bigg)\bigg(\frac{1}{\Omega B_{\rm eff}}\bigg),
    \label{eq:Power_spectrum_visibilities}
\end{equation}

where $\lambda$ is the centre wavelength of observing bandwidth, $k_B$ is the Boltzmann constant, $B_{\rm eff}$ is the effective bandwidth, $D_c$ is the comoving distance at the redshift of our measurement, $\Delta D_c$ is comoving distance parallel to the line of sight, $\Omega$ is the field of view solid angle, $\tilde{V}_b$ is the Fourier transformed visibility, and $b$ is the visibility baseline length.
We can map $\tau$ to the line-of-sight cosmological Fourier wavevector $k_{\parallel}$ using the relation \citep{Thyagarajan2013}:
\begin{equation}
    k_{||} = \frac{2\pi \nu_{21} H_0 E(z)}{c(1+z)^2} \, \tau,
\end{equation}
where $\nu_{21} = 1420$ MHz, $H_0$ is the Hubble constant, $E(z)=[\Omega_m(1+z)^3+\Omega_\Lambda]^{1/2}$, $\Omega_m$ is the normalized matter content and $\Omega_\Lambda$ is the normalized dark energy content.

\begin{figure*}
    \centering
    \includegraphics[scale=0.5]{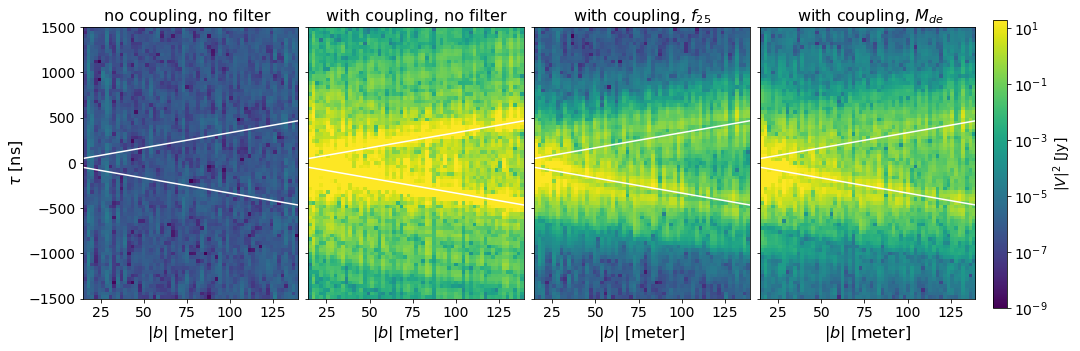}
    \caption{The residual of the squared visibilities in delay space as a function of baseline length at low band. The residuals are average over all baselines of equal length. The residual is taken with respect to the noise-free visibility model without mutual coupling. We show
residuals in the case where calibration assumes raw data visibility products without mutual coupling and is thus perfect, leaving only noise in the residual (first panel from the left) and raw
 data visibility with mutual coupling and no filter (second panel from left), a $f_{25}$ notch filtered scenario (third panel from the left) and lastly $M_{de}$ main lobe filtered scenario (right panel). The white line
marks the horizon limit of the baselines, which bounds the natural extent of foreground emission in the data.}
    \label{fig:wedge_residual filtered_25}
\end{figure*}

In Figure~\ref{fig:wedge_residual filtered_25}, we show the residual visibilities obtained after calibrating raw data with gains recovered from the data set with and without mutual coupling. The residual visibilities are obtained by taking the difference between calibrated visibilities and the sky model visibilities with mutual coupling effects. Recall that simulated true sky model visibilities contain effects of mutual coupling, thus a perfect calibration needs to yield calibrated visibilities that match closely with the sky model visibilities that contain effects of mutual coupling. As expected, when the raw data visibility products are without mutual coupling, we get a perfect calibration, yielding robust gains, and thus, the calibrated visibilities differ from the sky model visibilities by noise. Hence, the residual visibilities obtained are just noise (left panel). However, when our simulated raw data includes the effects of mutual coupling, our calibration is not perfect, and the recovered gains (Figure~\ref{fig:abscal_gains_non_redundant_no_filter_low}), which are subsequently used to calibrate the raw data are erroneous. Thus calibrated visibilities differ significantly from the true sky model visibilities. Additionally, when the recovered gains are used to calibrate the raw data because of their non-smooth spectral structure (due to calibration errors) they impart spectral structure on the relatively smooth foreground spectra, some of the foreground emission is pushed to higher delays beyond the horizon limit (white line). This is most prominent on the shorter baselines. In the panel second panel from the left, the residual visibilities have foreground emission at short baselines beyond the horizon limit, and this residual foreground emission extends up to delays of $\tau\sim 500 $~ns.  When the notch filter $f_{25}$ is applied (third panel from left) before calibration, more robust gains are used to calibrate the unfiltered raw data. We find that the calibrated visibilities closely match the sky model visibilities. As a result, the foreground residual emission beyond the horizon limit is significantly reduced by a factor of $\sim 10^2$. The excess residual foreground within the horizon line poses no issue, as this region in delay space is expected to be occupied by foreground emission. However, when the main lobe filter $M_{be}$ is used (right panel), there is comparatively less reduction in the foreground emission in residual emission compared to the notch filter. This is expected as the gains recovered when the main lobe filter is used are not as robust as gains obtained through the use of the notch filter, and can therefore cause foreground leakage when they are used to calibrate the data.  

\begin{figure*}
    \centering
    \begin{tabular}{cc}
       \includegraphics[scale=0.6]{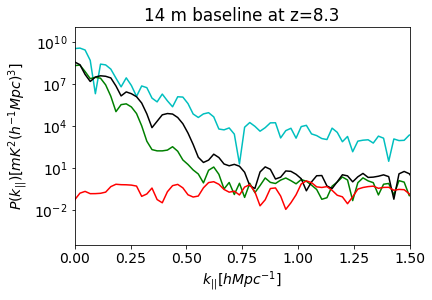}  & \includegraphics[scale=0.6]{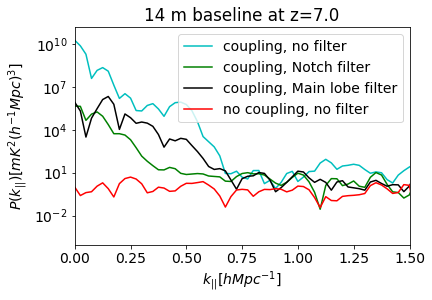} 
         
    \end{tabular}
    
    \caption{The residual delay power spectrum of the calibrated raw visibilities with respect to the sky model (noise-free) visibilities, averaged over 14~m redundant baselines and, over time, integrations, with LST range $[0^h-1^h]$. Left: Low band at redshift $z=8.3$, where raw data is without mutual coupling (red) and with mutual coupling (cyan), and then having applied the $f_{25}$ notch filter (green). Right: Same, but for the high sub-band at redshift $z=7.0$.}
    \label{fig:power_spectrum_low}
\end{figure*}

Figure~\ref{fig:power_spectrum_low} further shows the residual power spectrum for 14~m redundant baselines for both low and high sub-bands. The foreground residual power is significantly less in the high sub-band than in the low sub-band as the foreground residual emission is reduced to the noise level at $k_\parallel \sim 0.75$~h~Mpc$^{-1}$ in high sub-band, but at low sub-band, there is still residual power at $k_\parallel > 1.5$~h~Mpc$^{-1}$ (cyan line). This is again due to the increased mutual coupling effects at the low sub-band.  We also note the main-lobe filter also suppresses foreground residual emission (black line). However, the suppression is not as significant as the notch filter (green line) in both the low and high sub-band. In the low band, the notch filter suppresses residual foreground by a factor $\sim 10$ more than the main lobe filter at $k_\parallel < 0.75$~h~Mpc$^{-1}$. Additionally in the low sub-band, with the notch filter, the residual emission is reduced to levels below the noise floor at $ k_\parallel\approx 0.60$~h~Mpc$^{-1}$, whereas for the main lobe filter, this only occurs at $k_\parallel\approx 0.75$~h~Mpc$^{-1}$. In the high sub-band, the reduction of residual power by the use of the notch filter is by a factor of $\sim 10^3$ larger than that of the main lobe filter at $k_\parallel \approx 0.5$~h~Mpc$^{-1}$. Thus, the use of the $f_{25}$ notch filter produces robust gains that significantly reduce the foreground power leakage.

\begin{figure*}
    \centering
    \begin{tabular}{cc}
       \includegraphics[scale=0.6]{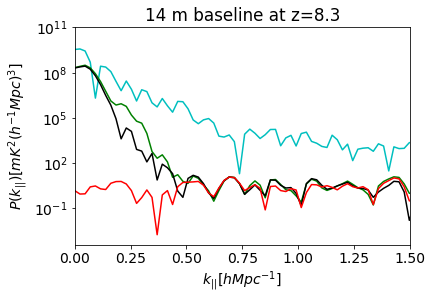}  & \includegraphics[scale=0.6]{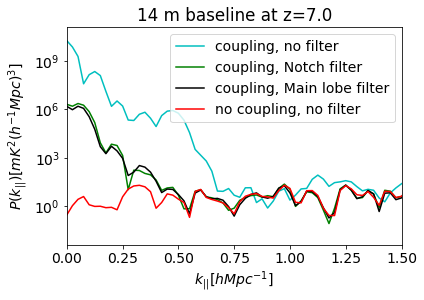} 
         
    \end{tabular}
    
    \caption{The residual delay power spectrum of the calibrated raw visibilities with respect to the sky model (noise-free) visibilities, averaged over 14~m redundant baselines and, over ten time integrations at LST centred at $2^h$. Left: Low band at redshift $z=8.3$, where raw data is without mutual coupling (red) and with mutual coupling (cyan), and then having applied the $f_{25}$ notch filter (green). Right: Same, but for the high sub-band at redshift $z=7.0$.}
    \label{fig:power_spectrum_low_at_2h}
\end{figure*}

Lastly, Figure~\ref{fig:power_spectrum_low_at_2h} shows the residual power from 14 m redundant baselines at an LST centred at $2^h$, typically used for calibration of HERA data \citep{Chapman2016}. The sky at this LST has strong compact sources and the diffuse component of the sky is colder compared to the LST of $0^h$. As a result, the error in the calibration is reduced significantly, as evident by the low recovered chi-square (Figure~\ref{fig:non_redundant_chis_low_as_function_LST}). We again note that in this field the use of fringe rate filters improves the calibration as the foreground residual power is reduced by $10^2$ at $k_\parallel\approx 0.5$~h~Mpc$^{-1}$. We, however, note that there is no significant difference in the improvement in calibration when either the main lobe or notch filter is used, particularly at the high sub-band. There is slightly better performance of the main lobe filter at low sub-band at $k_\parallel\approx 0.25$~h~Mpc$^{-1}$. In this field, the diffuse emission component falling on the main lobe is cold, compared to bright compact sources, however, the hot diffuse component still resides in the side lobes of the primary beam and thus would introduce calibration errors if the emission is not filtered out (cyan line). When this emission is filtered through with a notch filter (green line), the calibration is improved, and even more so when the main lobe is used, due to more of the emission being filtered. In summary, both filters have shown effectiveness at reducing foreground power leakage due to mutual coupling effects, with the notch filter showing better performance when the diffuse emission dominates the sky. Both filters were also shown to be effective at mitigating calibration due to the use of an incomplete sky model during calibration \citep{Charles2023}, especially the main lobe filter at LST of $2^h$. Thus, the main lobe filter can be used to mitigate both calibration errors due to incomplete sky modelling and the effects of mutual coupling.  

\section{Conclusion}
\label{sec:conclusion}

In this paper, we investigate the prospect of mitigating the effects of mutual coupling in calibration through the application of fringe rate filters prior to calibration. We make use of two types of filters; baseline dependent main lobe filter and baseline independent notch filter. The main lobe filter is designed to attenuate all emission that in not within the main lobe of the primary beam, whilst the notch filter is designed to attenuate emission that is near zero fringe rates. For a baseline in the HERA array with an East-West projection greater than 30~m, this emission will typically resides on the sidelobes of the primary beam. 

We show that mutual coupling effects lead to non-smooth calibration errors, which impart spectral structure in gains derived from calibration in the delay $\tau >250$~ns. When these gains are used to calibrate data, we find significant contamination of the EoR window by foreground emission, which is about 4 orders of magnitude at $k_\parallel \approx 0.5$~h~Mpc$^{-1}$ at $z=8.3$ for a 14 m redundant baselines. By applying the filters before calibration we find in this paper that the calibration significantly improves, as the recovered gains are much more robust i.e. much closer to the true gains, and when these gains are used to calibrate data we observe a significant decline in foreground contamination of EoR window. The foreground contamination is reduced by at least 2 orders of magnitude at  $k_\parallel \approx 0.5$~h~Mpc$^{-1}$. In addition, we find that the application of the notch filter leads to reduced calibration errors compared to the main lobe filter. We find that the foreground contamination is an order of magnitude below for notch filter compared to the main lobe filter. This is mainly due to mutual coupling effects that are within the main lobe of the primary beam, to which the main lobe filter retains, thus resulting in erroneous gains compared to the notch filter, where a portion of this emission is filtered.

Although the use of fringe rate filters has shown significant improvement in the calibration, we, however, note that more investigation is still needed, because real observations normally include flags, that can potentially affect the filtering of data. When data containing flags is Fourier transformed along the time axis, we expect there to be ``spectral leakage” resulting in the mixing of fringe rate modes of the sky due to the sidelobe structure introduced by the null/flags in the data. Thus, flagging on the time axis can reduce the effectiveness of the filter. To reduce the impact of missing data points, the data can be inpainted prior to applying the filters thus preventing the mixing of fringe rate modes of sky prior to filtering the data. A variety of inpainting techniques for radio data could be leveraged in this effort \citep[e.g.,][]{Parsons2009, Kern2021, Kennedy2022, Pagano2023}.

\section*{Data Availability}
Data used in this work may be made available upon reasonable request to the corresponding author. The code used for the analysis of this work is publicly available at \href{https://github.com/Ntsikelelo-Charles/Fringe_rate_filters}{https://github.com/Ntsikelelo-Charles/Fringe\_rate\_filters}. This work relied on publicly available and open-sourced community Python software, including {\sc numpy}\citep{2020NumPy-Array}. Visibility simulations were made using {\sc matvis} \citep{Kittiwisit2023}, and simulation of mutual coupling effects {\sc hera\_sim} (\href{https://github.com/HERA-Team/hera_sim}{https://github.com/HERA-Team/hera\_sim}) and {\sc pyuvdata} \citep{Hazelton2017}. 

\section*{Acknowledgements}

This work was supported in part by the Italian Ministry of Foreign Affairs and International Cooperation”, grant number ZA23GR03. NK acknowledges support from NASA through the NASA Hubble Fellowship grant \#HST-HF2-51533.001-A awarded by the Space Telescope Science Institute, which is operated by the Association of Universities for Research in Astronomy, Incorporated, under NASA contract NAS5-26555. MGS acknowledges support from the South African Radio Astronomy Observatory and National Research Foundation (Grant No. 84156). OMS's research is supported by the South African Research Chairs Initiative of the Department of Science and Technology and National Research Foundation (grant No. 81737). AL and RP acknowledge support from the Trottier Space Institute, the Canadian Institute for Advanced Research (CIFAR) Azrieli Global Scholars program, a Natural Sciences and Engineering Research Council of Canada (NSERC) Discovery Grant, a NSERC/Fonds de recherche du Québec-Nature et Technologies NOVA grant, the Canada 150 Research Chairs Program, the Sloan Research Fellowship, and the William Dawson Scholarship at McGill.



\bibliographystyle{mnras}
\bibliography{main} 




\bsp	
\label{lastpage}
\end{document}